\documentclass[%
superscriptaddress,
nofootinbib,
amsmath,amssymb,
aps,
pra,
longbibliography,
showkeys,
notitlepage
]{revtex4-1}

\usepackage[left=2.5cm,top=3cm,right=2.5cm,bottom=3cm,bindingoffset=0.5cm]{geometry}
\usepackage{graphicx}
\usepackage{hyperref}
\usepackage[bitstream-charter, greekfamily=default]{mathdesign}
\usepackage{lipsum}
\usepackage{xcolor}
\usepackage{subeqnarray}

\newcommand{\betaf}{\mathrm{B}}

\newcommand{\dint}{\mathrm{d}}

\newcommand{\abs}[1]{\left\vert #1 \right \vert}
\renewcommand{\epsilon}{\varepsilon}

\begin{document}

\title{Schr\"odinger's ants: A continuous description of Kirman's recruitment model} 

\author{Jos\'{e} Moran}
\email{email: jose.moran@polytechnique.org}
\affiliation{Centre d'Analyse et de Math\'{e}matique Sociales, EHESS, 54 Boulevard Raspail, 75006 Paris}
\affiliation{Chair of Econophysics and Complex Systems, Ecole polytechnique, 91128 Palaiseau Cedex, France}

\author{Antoine Fosset}%
\affiliation{Chair of Econophysics and Complex Systems, Ecole polytechnique, 91128 Palaiseau Cedex, France}
\affiliation{LadHyX UMR CNRS 7646, Ecole polytechnique, 91128 Palaiseau Cedex, France}

\author{Michael Benzaquen}%
\affiliation{Chair of Econophysics and Complex Systems, Ecole polytechnique, 91128 Palaiseau Cedex, France}
\affiliation{LadHyX UMR CNRS 7646, Ecole polytechnique, 91128 Palaiseau Cedex, France}
\affiliation{Capital Fund Management, 23 Rue de l'Universit\'{e}, 75007 Paris\medskip}

\author{Jean-Philippe Bouchaud}
\affiliation{Capital Fund Management, 23 Rue de l'Universit\'{e}, 75007 Paris\medskip}
\affiliation{Chair of Econophysics and Complex Systems, Ecole polytechnique, 91128 Palaiseau Cedex, France}

\date{\today}

\begin{abstract}
We show how the approach to equilibrium in Kirman's ants model can be fully characterized in terms of the spectrum of a Schr\"odinger equation with a P\"oschl-Teller ($\tan^2$) potential. Among other interesting properties, we have found that in the bimodal phase where ants visit mostly one food site at a time, the switch time between the two sources only depends on the ``spontaneous conversion'' rate and {\it not} on the recruitment rate. More complicated correlation functions can be computed exactly, and involve higher and higher eigenvalues and eigenfunctions of the Schr\"odinger operator, which can be expressed in terms of hypergeometric functions.
\end{abstract}

\maketitle


\section*{Introduction}

Kirman's \emph{ant model}~\cite{kirman1993ants} undoubtedly stands among some of the most inspiring toy models in the behavioral economics literature. While initially inspired by the experiment described below, its conclusions have  implications much beyond collective animal behaviour, and has been used to model shifts in sentiment of economic agents, trend reversal in financial markets, herding and social influence, etc. Kirman's model is also akin to another famous model in population dynamics with competing species: the Moran model \cite{moran_1958}. \\

Several decades ago entomologists were puzzled by the following observation \cite{deneubourg1990self, beckers1990collective}. Ants, faced with two identical and inexhaustible food sources $A$ and $B$, tend to concentrate on one of them for a while, but occasionally switch to the other. Such intermittent herding behavior is also observed in humans choosing between equivalent restaurants~\cite{becker1991note}, or in financial markets~\cite{scharfstein1990herd,shiller1986survey,lux1995herd} consistent with large endogenous fluctuations. Clearly the asymmetric exploitation observed in ants does not seem to correspond to the equilibrium state of an isolated representative ant with rational expectations. The phenomenon is rather to be explained in terms of interactions between individual ants, or, as put by biologists, recruitment dynamics.  To account for such intricate  behavior, Kirman proposed a simple and insightful model~\cite{kirman1993ants} based on tandem recruitment that we now recall.\\

Consider $N$ ants and denote by $k(t)\in[0,N]$ the number of ants feeding on source $A$ at time $t$. When two ants meet, one of them converts the other with probability $\mu/N$, but each ant may in addition change its own mind spontaneously with probability $\epsilon$. Within such a simple setting, Kirman was able to show that, in the large $N$ limit, the stationary state depends only on a parameter $\alpha := \epsilon/\mu$. When $\alpha >1$ the distribution is unimodal, with a maximum at $k=N/2$, whereas for $\alpha < 1$ the stationary distribution of $k$ is bimodal, with maximum probability for $k=0$ and $k=N$ (corresponding to the situation observed in the experiments). 
It is remarkable that the interesting $\alpha<1$ regime can be obtained even for weakly persuasive agents ($\mu$ small) provided self-conversion $\epsilon$ is itself low enough.\\

The most important point is that in the $\alpha <1$ regime no one of the $k$ states is, in itself, an equilibrium. Although the system can spend a long time at $k=0,N$ (local stationarity) these states cannot be considered as such: all the states are always revisited and there is no convergence to any particular state, discarding also the notion of multiple equilibria. Rather, there is perpetual change, and the system's natural endogenous dynamics is only in a \emph{statistical equilibrium}. Most economic models focus on finding the equilibrium to which the system will finally converge, and the system may only be knocked off its path by large exogenous shocks.  

Yet financial markets, and even larger economic and social systems, display a number of regular large switches (correlations, mood of investors etc.) which do not seem to be always driven by exogenous shocks. In Kirman's stylised setting such switches can be understood endogenously. Several extensions of the model have been proposed~\cite{lux1995herd,kirman2002microeconomic,gilli2003global}. 
In particular, the original version of Kirman's model does take into account the heterogeneity in encounter probabilities induced by the topology of the social network; but one can easily (at least numerically!) modulate the probability of encounters according to their distance along such a network, for example restricting recruitments to nearest neighbours only. 
\\

In the present paper we present a continuous description of Kirman's ant model which notably allows us to derive the typical switching time, using classical methods from statistical physics and quantum mechanics. 

\section{Master Equation}

As mentioned above the original model describes $N$ ants faced with two identical food sources, with the relevant dynamical variable being $k$, the number of ants feeding  on -- say -- source A. Each time step allows an ant to either switch randomly to the other food source with probability proportional to $\epsilon$, or get recruited by another ant from the other food source with probability proportional to $\mu$. \\

Defining the unit of time as the time required for all the ants to make a decision, leads to $\mathrm dt = 1/N$ as the infinitesimal time step. It is also clear that, to remain intensive in the large $N$ limit, the probability to interact with another ant should be proportional to $1/N$. Altogether, we may write a Master equation for the evolution of the probability $\mathbf P(k,t)$ that there are $k$ ants feeding at source A at time $t$:
\begin{eqnarray}\label{eq:meq_pk}
\mathbf P\left(k, t+\frac{1}{N}\right)- \mathbf P(k,t) &=& \frac{1}{N}\Big\{W(k+1 \to k)\mathbf P(k+1,t) + W(k-1\to k)\mathbf \mathbf P(k-1,t)\nonumber \\
&& - \big[ W(k\to k-1) - W(k\to k+1)\big]\mathbf P(k,t)\Big\},
\end{eqnarray}
where the transition rates are given by:
\begin{subeqnarray}\label{eq:tran_rates}
W(k\to k+1) &=& \left(1 - \frac{k}{N}\right)\left(\epsilon+\frac{\mu}{N}\frac{k}{N-1}\right)\\
W(k\to k-1) &=& \frac{k}{N}\left(\varepsilon + \frac{\mu}{N}\frac{N-k}{N-1}\right).
\end{subeqnarray}
Note that this specification only differs from Kirman's original one in the rescaling of the recruitment rate by $N$. With the notations of \cite{kirman1993ants}, $1-\delta = \mu/N$.

\section{Continuous description and Fokker-Planck equation}

Here we follow Kirman's original paper \cite{kirman1993ants} and derive a proper continuous-time Fokker-Planck equation in the limit $N\to\infty$.\\

We define the variable $x=\frac{k}{N}\in[0;1]$ together with its probability density function $f(x,t)$. Taking the continuous limit $N\to\infty$ of Eq.~\eqref{eq:meq_pk} leads to the following Fokker-Planck equation~\cite{risken1996fokker}:
\begin{equation}\label{eq:f_fp}
\partial_t f = \partial_x J^f , \quad \text{with} \quad J^f(x,t) = -\varepsilon (1-2x)f(x,t) + \mu \partial_{x}\left[x(1-x)f(x,t)\right],
\end{equation}
the probability flux, see Appendix~\ref{sec:FPdetails} for the details of the calculations and the first $1/N$ corrections. The conservation of the number of ants in the model is ensured by the condition $J^f(x,t)=0$ at the boundaries $x=0$ and $x=1$ at all times. Equation~\eqref{eq:f_fp} corresponds to the following stochastic process for $x$:
\begin{equation}\label{eq:sde_x}
\dot x = \varepsilon(1-2x) + \sqrt{2\mu x(1-x)}\eta(t)\, ,
\end{equation}
with $\eta$ a Gaussian white noise with unit variance. One can note that while the drift term $\varepsilon(1-2x)$ is maximal at the boundaries and tends to pull $x$ towards $1/2$, the noise term has the opposite effect.  The diffusion constant is proportional to $\sqrt{2\mu x(1-x)}$ and is maximal at $x=1/2$ and so tends to push the system away from $x=1/2$. Note that this stochastic process is very similar to the Moran model of genetic population dynamics \cite{moran_1958} -- with the same diffusion term $\propto \sqrt{x(1-x)}$ -- where the analogue of the number of ants at each food source is the proportion of genes from two competing alleles (A or B) \cite{wright1942}. The $\varepsilon$ term corresponds to spontaneous mutations. When $\epsilon = 0$, there is a non zero probability that the whole population becomes of type A or B after a finite time, corresponding to $\delta(x)$ or $\delta(1-x)$ contributions to $f(x,t)$ with a time dependent weight, see \cite{Kimura1955}, and \cite{McKane2007} for a recent thorough discussion.\\

When $\varepsilon > 0$, one can check that the normalised stationary distribution $f_0(x)$, obtained by setting $J^f(x,t)=0$, writes:
\begin{equation}\label{eq:stat_sol}
f_0(x)=\frac{\Gamma(2\alpha)}{\Gamma^2(\alpha)}\big[x(1-x)\big]^{\alpha - 1},\quad \text{with}\quad \alpha:= \frac{\varepsilon}{\mu}\, .
\end{equation}
This result is the same as that obtained by F\"ollmer and Kirman in~\cite{kirman1993ants}. \\

\begin{figure}[tb]
    \centering
    \includegraphics[width = 1\textwidth]{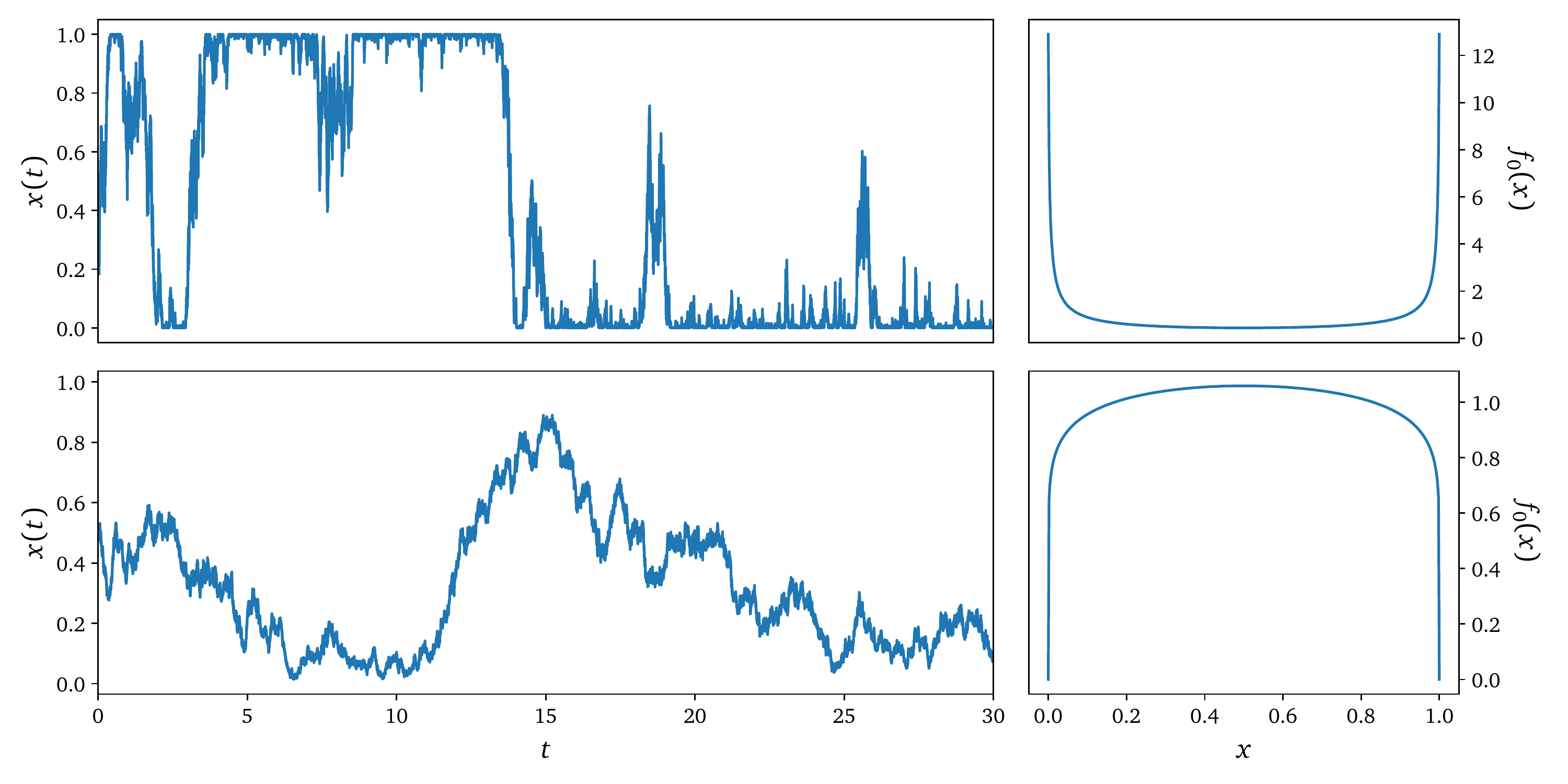}
    \caption{Simulations of the model in the continuous limit. The top plots correspond to $\alpha = 0.1<1$ while the bottom ones to $\alpha = 2>1$. Both simulations were run with $\epsilon = 0.1$. The left panels display the evolution of $x(t)$ as defined in Eq.~\eqref{eq:sde_x}.  The right panels display the corresponding stationary probability densities, as given by Eq.~\eqref{eq:stat_sol}.}
    \label{fig:model_plots}
\end{figure}

Upon looking at the behaviour of the solution, shown in Fig.~\ref{fig:model_plots}, one can see that there is a clear  transition in the behaviour of the model at $\alpha_c = 1$. For $\alpha > \alpha_c$, the stationary density in Eq.~\eqref{eq:stat_sol} is maximal at $x=1/2$, and the dynamics shows that $x(t)$ fluctuates around $1/2$, corresponding to a situation where the ants are, on average, evenly distributed across both food sources. 
For $\alpha < \alpha_c$ the density $f_0$ diverges at the boundaries. The top left panel in Fig.~\ref{fig:model_plots} shows that this corresponds to a very different picture, in which nearly all of the ants choose either one of the sources for a certain amount of time, until a noise-induced ``avalanche'' causes a switch over to the other source. It is also easy to check that in the absence of noise (and $\alpha\to 0$) the long-time stationary density is given by  $f_0(x)=\frac{1}{2}\left[\delta(x)+\delta\left(x-{1}\right)\right]$, a situation discussed at length in \cite{McKane2007}.\\

Having this in mind, a natural question to ask is: Given a certain initial condition $f(x,0)=\delta(x-x_0)$, how long does it take for the system to converge to the stationary state, or equivalently, how long does it take for the ants to switch from one source to the other in the $\alpha<1$ regime?

\section{Schr\"odinger's equation and general solution}

Here we obtain a full dynamical solution in terms of the eigenvalues and eigenfunctions of a certain quantum mechanical Hamiltonian.\\

Using the It\^o rule~\cite{ito1951stochastic}, one can see that introducing a change of variables $\varphi(x)$ in Eq.~\eqref{eq:sde_x} yields a noise term proportional to $\sqrt{x(1-x)}\varphi'(x)$, and so motivates a choice satisfying $\varphi'(x)=1/\sqrt{x(1-x)}$. We therefore define a new, more convenient,  variable $\varphi \in[-\pi/2,\pi/2]$ as:
\begin{equation}\label{eq:phi_def}
\sin \varphi =2x-1.
\end{equation}
 The corresponding Fokker-Planck equation for its probability density $g(\varphi,t)$ writes:
\begin{equation}\label{eq:phi_fp}
\partial_t g = \mu \partial_\varphi J^g, \quad \text{with} \quad
J^g(\varphi,t) = 2\beta \tan\varphi g(\varphi,t)+\partial_{\varphi}g(\varphi,t)\, ,\quad \text{and} \quad \beta := \alpha - \frac{1}{2}\, ,
\end{equation}
where the probability flux must now verify $J^g(\pm \pi/2,t)=0$ at all times. Setting again $J^g=0$ everywhere, one  finds the normalized stationary solution:
\begin{equation}\label{eq:gstat}
g_0(\varphi)=\frac{\Gamma\left(\alpha + \frac{1}{2}\right)}{\sqrt{\pi}\Gamma(\alpha)}(\cos\varphi)^{2\alpha -1}.
\end{equation}
The advantage of this formulation in $\varphi$ is that, in contrast with the former, the second order derivative term $\partial_{\varphi\varphi}$ in Eq.~\eqref{eq:phi_fp} only depends on $\varphi$ through $g(\varphi,t)$. Standard techniques for the resolution of Fokker-Plank equations, see e.g.~\cite{risken1996fokker}, motivate the introduction of a function $\Psi$ such that:
\begin{equation}\label{eq:psi_def}
g(\varphi,t) := \sqrt{g_0(\varphi)}\Psi(\varphi,t)\,,
\end{equation}
and $\Psi(\varphi,t) \to \sqrt{g_0(\varphi)}$ when  $t\to\infty$.\\

Combining Eqs.~\eqref{eq:phi_fp} and~\eqref{eq:psi_def}  one obtains a Schr\"odinger-like equation of the form~\cite{cohen2006quantum}:
\begin{equation}\label{eq:schrodinger_psi}
-\frac{1}{\mu} \partial_t \Psi = \mathbf{H}\Psi,
\end{equation}
where the Hamiltonian $\mathbf{H}$ is defined as:
\begin{equation}\label{eq:schrodinger_psi2}
\mathbf{H} := - \partial_{\varphi\varphi} + V(\varphi), \quad V(\varphi):=-\beta+\beta(\beta-1)\tan^2\varphi,
\end{equation}
and with boundary conditions given by:  
\begin{equation}\label{eq:schrodinger_bc}
\left[\cos^\beta\hspace{-0.5mm} \varphi \big( \beta \tan \varphi \Psi(\varphi,t)+\partial_{\varphi}\Psi(\varphi,t)\big) \right]_{\varphi = \pm \pi/2} =0\, .
\end{equation}
We have left the $\mu$ parameter out of the Hamiltonian $\mathbf H$ in order to ease the comparison to the canonical form presented in~\cite{Nieto,tacseli2003exact}. The $\tan^2$ term in Eq.~\eqref{eq:schrodinger_psi} is known as the P\"oschl-Teller potential~\cite{Nieto}, which was fully solved in the case $\beta>0$ with boundary conditions $\Psi(\pm\pi/2,t)=0$ in \cite{tacseli2003exact}. To be applicable to our framework, we shall verify that their solutions also satisfy Eq.~\eqref{eq:schrodinger_bc} in the general case $\beta>-1/2$.
The Hamiltonian $\mathbf{H}$ is Hermitian (contrarily to the Fokker-Planck operator) and has a discrete set of orthogonal eigenfunctions and eigenvalues, given by:
\begin{equation}\label{eq:hamiltonian_diag}
\mathbf{H}\Psi_n = {\mathcal E}_n \Psi_n\, ,
\end{equation} 
where, splitting into even ($n=2k$) and odd ($n=2k+1$) states:
\begin{subeqnarray}\label{eq:sol_array}
        {\mathcal E}_{n} &=& n(2\alpha + n-1) \, ,\\
         \Psi_{2k}(\varphi) &=& A_{2k}(\beta)~_2F_1\left(-k,\beta+k;\beta+\textstyle \frac{1}{2}, \cos^2 \varphi \right)\cos^\beta \varphi ,\\
        \Psi_{2k+1}(\varphi) &=&  A_{2k+1}(\beta)~_2F_1\left(-k,\beta+k+1;\beta+\textstyle  \frac{1}{2}, \cos^2 \varphi\right) \sin \varphi~ \cos^\beta \varphi,
\label{eq:n_odd}
\end{subeqnarray}
 with $_2F_1$  the ordinary hypergeometric function.\footnote{Here, the function $_2F_1$ takes the form of a polynomial: $\displaystyle 
~_2F_1\left(-k,a;b, u\right)=\sum_{l=0}^{k} \binom{k}{\ell}(-1)^\ell \frac{\Gamma(a+\ell)}{\Gamma(a)}\frac{\Gamma(b)}{\Gamma(b+\ell)}u^\ell$ for any integer $k$.
}
The coefficients $A_{n}$ are set such as  to ensure normalisation, $\int_{[-\pi/2,\pi/2]}\Psi_n\Psi_m = \delta_{n,m}$, and can be expressed as integrals of  hypergeometric functions. Note that  the parity of $n$ also defines the parity of the function $\Psi_n$ with respect to the $y$-axis. One can then easily check that for all $n$ (both even and odd):
\begin{equation}\label{eq:equivalence}
\cos^\beta\hspace{-0.5mm} \varphi\left[\beta \tan \varphi\Psi_n(\varphi)+ \Psi'_n(\varphi)\right] \underset{\varphi\to\pm \pi/2}{\sim}\left(\frac{\pi}{2}\mp \varphi\right)^{1+2\beta},
\end{equation}
which, since $\beta>-1/2$, ensure that the boundary conditions  given by Eq.~\eqref{eq:schrodinger_bc} are satisfied.
Noting that $\Psi_0 = \sqrt{g_0}$, the general solution of Eq.~\eqref{eq:schrodinger_psi} then reads:
\begin{equation}\label{eq:schrodinger_sol}
\Psi(\varphi,t) = \lambda_0\sqrt{g_0(\varphi)}+\sum_{n>1}\lambda_n \Psi_n(\varphi)e^{- \mu {\mathcal E}_n t},
\end{equation}
with $\lambda_n $ given by the projections of the initial conditions on each mode $n$, namely $\lambda_n  = \int_{-\pi/2}^{\pi/2}\dint \varphi~\Psi_n(\varphi)\Psi(\varphi,0)$. \\

Back to the physical variable $x$, the initial condition $f(x,0)=\delta(x-x_0)$ becomes $g(\varphi,0)=\delta(\varphi-\varphi_0)$ with $\varphi_0 =\arcsin\left(2x_0-1\right)$. Further using~Eq.~\eqref{eq:psi_def}, it is easy to see that the initial  condition in turn translates into
$
\Psi(\varphi,0)={\delta\left(\varphi-\varphi_0\right)}/{\sqrt{g_0(\varphi)}}
$. The full solution for $g(\varphi,t)$ follows:
\begin{equation}\label{eq:g_sol}
g(\varphi,t) = g_0(\varphi)+\sum_{n>1}e^{-\mu{\mathcal E}_n t}\Psi_n(\varphi_0)\sqrt{g_0(\varphi)}\Psi_n(\varphi)\, ,
\end{equation}
with the orthogonality between $\Psi_0=\sqrt{g_0}$ and $\Psi_n$ for $n>1$ ensuring that ${\int_{-\pi/2}^{\pi/2}\dint \varphi~g(\varphi,t)=\int_{-\pi/2}^{\pi/2}\dint \varphi~g_0(\varphi) =1}$, or equivalently for $f(x,t)$:
\begin{equation}\label{eq:f_sol}
f(x,t) = f_0(x) + \sum_{n>1} e^{-\mu {\mathcal E}_n t}\Psi_n(\varphi_0)f_n(x)\, ,
\end{equation}
with: 
\begin{equation}
    f_n(x) = \frac{\sqrt{g_0(\varphi(x))}\, \Psi_n(\varphi(x))}{2\sqrt{x(1-x)}}\, ,
\end{equation}
(see Appendix~\ref{sub:explicit_expressions} for an explicit expression). Equation~\eqref{eq:f_sol} is the central result of the present communication.

\begin{figure}[tb]
    \centering
    \includegraphics[width=\textwidth]{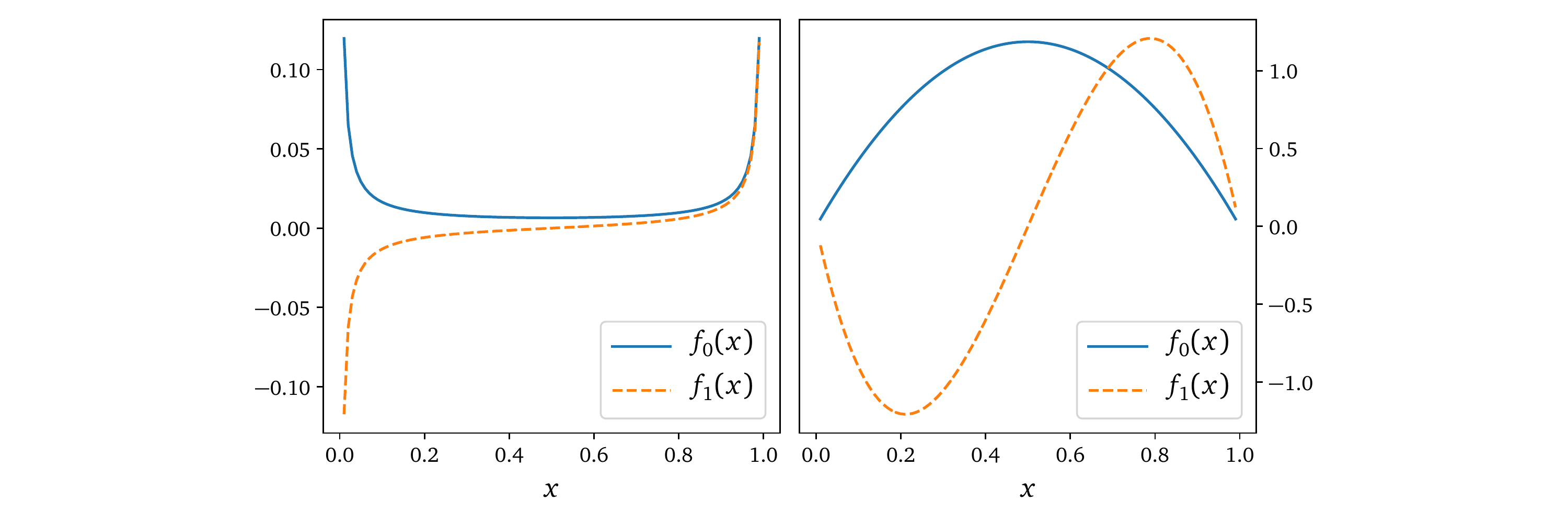}
    \caption{A plot showing the shape of the first two modes $f_0(x)$  and $f_1(x)$. The top panel corresponds to $\alpha=0.1$, while the one on the bottom corresponds to $\alpha = 2$. $f_0(x)$ is the stationary state, whereas $f_1(x)$ is the slowest decaying mode, that corresponds to hopping between the two food sources. }
    \label{fig:f_fig}
\end{figure}

\section{Relaxation towards the stationary state}

With the full dynamical solution of Eq.~\eqref{eq:f_sol} at hand, one can see how long a system initially prepared at an initial value $x_0\approx 0$, for example, takes to explore the whole space. In other words, one can ask how much time $\tau$ is required to reach, say, $x(\tau)\approx 1$ with a reasonable probability. 

Since the stationary distribution $f_0$ has weight on the whole interval $[0;1]$, this time $\tau$ is none other than the relaxation time (or ergodic time) $\tau_{\text{R}}$ required to converge to stationarity. Owing to the form of Eq.~\eqref{eq:f_sol} this convergence is asymptotically exponential, with the slowest mode given by $n=1$. Hence, we find:
\begin{equation}\label{eq:tau_rel}
\tau_{\text{R}} := \frac{1}{\mu{\mathcal E}_1} \equiv \frac{1}{2\epsilon}.
\end{equation}
Perhaps surprisingly, this relaxation time depends only on the noise intensity $\varepsilon$, but not on the recruitment intensity $\mu$. Since $n=1$ corresponds to the slowest mode of the system, it also governs the collective ``switch time'' between the two food sources, A and B -- see Fig. \ref{fig:f_fig}. 

\begin{figure}[tb]
    \centering
    \includegraphics[width=\textwidth]{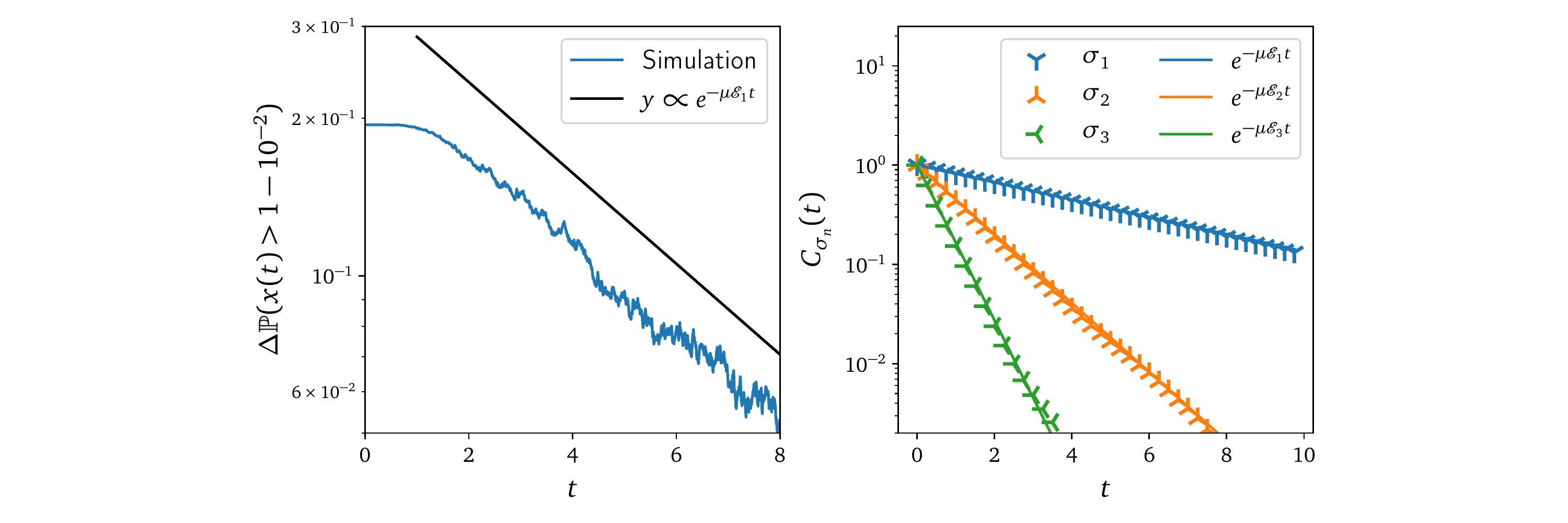}
    \caption{Left: plot of $\Delta\mathbb{P}(x(t)>1-10^{-2})$, defined as the difference between $\mathbb{P}(x(t)>1-10^{-2})$ and its stationary value, for $\varepsilon = 0.1$ and $\mu=0.5$. The initial condition is $x_0=10^{-2}$. Right: plots of the covariances $C_{\sigma_n}(t)=\text{Cov}\left[\sigma_n(x(T+t)),\sigma_n(x(T))\right]$, computed over simulations with $\varepsilon=0.1$ and $\mu=0.2$. The agreement with theoretical predictions is excellent.}
    \label{fig:switch_correls}
\end{figure}

We have checked our prediction for the switching time numerically by running trajectories starting at $x_0=\Delta x\ll 1$ and computing the probability $\mathbb{P}(x(t)>1-\Delta x)$. This quantity should converge to $\int_{[1-\Delta x;1]}f_0$ at an exponential rate $\propto e^{-\mu \mathcal{E}_1t}$, which is in perfect agreement with our simulations, see Figure~\ref{fig:switch_correls}. 

Similarly, given an initial condition $x_0=1/2$ where the ants are initially distributed evenly between the two sources, one may ask how long it takes for all the ants to ``decide'' on concentrating on one of them. Since this condition is equivalent to $\varphi_0=0$, and since $\Psi_1$ is an odd function of $\varphi$, it follows that $\Psi_1(\varphi_0)=0$ in this case. The convergence to the stationary distribution is then controlled by the second mode, with a much shorter relaxation time given by:
\begin{equation}\label{eq:tau_rel_bis}
\tau_{\text{R}}':=\frac{1}{\mu {\mathcal E}_2} \equiv \frac{1}{4\varepsilon+2\mu}.
\end{equation}

Directly applying tools from stochastic calculus on Eq.~\eqref{eq:sde_x}, one can obtain the following correlation functions (see Appendix~\ref{sec:stoch}):
\begin{equation}\label{eq:stoch}
\text{Cov}\left[\sigma_n(x(T+t)),\sigma_n(x(T))\right] \propto e^{-\mu \mathcal{E}_n t},
\end{equation}
where $\sigma_n(x)$ are polynomials of degree $n$ that allow one to ``diagonalize'' the evolution of the correlations:
\begin{equation}
\begin{split}
&\sigma_1(x)=x,\\
&\sigma_2(x)=x(1-x),\\ 
&\sigma_3(x)=(2x-1)\left[\left(1+\frac{2\alpha}{3}\right)(2x-1)^2-1\right].
\end{split}
\end{equation}
See Appendix~\ref{sec:stoch} for further details and Figure~\ref{fig:switch_correls} for a comparison with numerical results.

This result actually hides a deeper interpretation of the different modes $f_n$. In the case described above, one can surmise that the dynamics of the moments $\mathbb{E}[x]$, $\mathbb{E}[x^2]$ and $\mathbb{E}[x^3]$ are determined exclusively by the modes $f_1$,$f_2$ and $f_3$. In fact, focusing on any moment $\mathbb{E}\left[x^m\right]$, it is possible to prove that:
\begin{equation}\label{eq:moments}
\forall n>m,\quad B_{n,m}=\int_{0}^{1}\dint x~f_n(x)x^m = 0\, ,
\end{equation}
as well as for all values $n$ that do not have the same parity as $m$. This implies in fact that the dynamics of the moments $\mathbb{E}\left[x^m\right]$ are fully described by the modes $(f_1,\ldots,f_m)$, with only even values of $n$ contributing to even moments $m$ and vice-versa.
For example, for $m=3$ with the initial condition $x(0)=x_0$ we can compute:
\begin{equation}\label{eq:m3}
\mathbb{E}[x^3(t)] = B_{1,3}\Psi_1(\varphi_0)e^{-2\epsilon t}+B_{3,3}\Psi_3(\varphi_0)e^{-3(2\epsilon+2\mu)t},
\end{equation}
where the exact expression of $B_{n,m}$ is given in Appendix~\ref{ssec:moments}, Eqs.~\eqref{eq:momentsf} and \eqref{eq:inm}. Mind that $B_{0,m}$ is the stationary value of moment $\mathbb{E}[x^m(t)]$ for all moments.

\section{Conclusion}

In this work, we have shown how that the approach to equilibrium in Kirman's ants model can be fully characterized in terms of the spectrum of relaxation times, itself computable as the eigenvalues of a Schr\"odinger equation with a P\"oschl-Teller ($\tan^2$) potential. Note that similar techniques have been  recently applied to discuss the dynamics of wealth inequality in Ref. \cite{Gabaix_Lasry_Lions_Moll_2016}. Among other interesting properties, we have found that in the bimodal phase where ants visit mostly one food site at a time, the switch time between the two sources only depends on the ``spontaneous conversion'' rate $\epsilon$ and {\it not} on the recruitment rate $\mu$. This means that a single ant deciding on its own to explore an alternative food source can trigger an ``avalanche'' where the whole colony follows suit. More complicated correlation functions can be computed exactly, and involve higher and higher eigenvalues and eigenfunctions of the Schr\"odinger operator.\\

The possibility to solve exactly the dynamics of Kirman's model is of course intellectually satisfying. It is also important in view of the number of possible applications of such a model, recalled in the introduction, and which has reappeared recently in the context of self-fulfilling prophecies in a simple economic model \cite{bouchaud_farmer} and in the empirical study of the dynamics of fishers seeking to exploit fishing zones with finite resources \cite{moran_fish}. Our analytical approach can also be easily generalized to other models of genetic population dynamics, such as the general setting discussed in \cite{McKane2007}, as the change of variable we introduce always leads to a Schr\"odinger equation with a trigonometric potential provided the drift is linear in $x$. These equations may then be solved using known analytical tools \cite{Ciftci2013}.\\

We warmly thank Roger Farmer, Alan Kirman and Joachim Krug for fruitful discussions. This research was conducted within the \emph{Econophysics \& Complex Systems Research Chair} under the aegis of the Fondation du Risque, a joint initiative by the Fondation de l'\'Ecole polytechnique, l'\'Ecole polytechnique and Capital Fund Management.

\bibliography{biblio}

\clearpage

\appendix

\section{Derivation of the Fokker-Planck equation and stationary solution} \label{sec:FPdetails}

We define the continuous distribution $f(x,t)$ as:
\begin{equation}\label{eq:f_def}
f(x,t) = \lim_{N\to\infty}\sum_{k=0}^N \delta\left(x-\frac{k}{N}\right)\mathbf{P}(k,t)\, ,
\end{equation}
which amounts to replacing $\frac{k}{N}$ by $x$ in Eqs.~\eqref{eq:meq_pk} and \eqref{eq:tran_rates}. In this case, and to leading order in $\frac{1}{N}$, the term e.g. $W(k+1\rightarrow k)\mathbf{P}(k+1,t)$ reads:
\begin{equation}
     \left(1-\left(x+\frac{1}{N}\right)\right)\left(\varepsilon+\frac{\mu}{N}\left(x+\frac{1}{N} \right)\right)f\left(x+\frac{1}{N},t\right).
 \end{equation} 
We proceed similarly for all terms in the right-hand side of Eq.~\eqref{eq:meq_pk}, and Taylor-expand the left-hand side to leading order in the time variable, to obtain:
\begin{equation}
    \begin{split}{\partial_t f(x,t)} =& \textstyle\frac{\varepsilon}{\Delta} \big[ (x+\Delta)f(x+\Delta,t) - xf(x,t) - (1-x)f(x,t) + \left(1-(x-\Delta)\right)f(x-\Delta,t) \big]\\ & + \textstyle\frac{\mu}{\Delta^2} \big[ (x+\Delta)\left(1-(x+\Delta)\right)f(x+\Delta,t) +(x-\Delta)\left(1-(x-\Delta)\right)f(x-\Delta,t) - 2x(1-x)f(x,t) \big]\, ,
\end{split}
\end{equation}
where $\Delta = \frac{1}{N}$ for simplicity. We next Taylor-expand the right-hand side terms, such as e.g. $(x+\Delta)f(x+\Delta,t)\approx xf(x,t)+\Delta \partial_x\left[xf(x,t)\right]+\mathcal{O}(\Delta^2)$, to order $\Delta$ for the terms with prefactor $\epsilon/\Delta$ and to order $\Delta^2$ for the terms with prefactor $\mu/\Delta^2$. Gathering everything, we obtain the Fokker-Planck equation:
\begin{equation}\label{eq:appendix_fp}
    \partial_t f(x,t) = -\epsilon\partial_x\left[(1-2x)f(x,t)\right]+\mu \partial_{xx}\left[x(1-x)f(x,t)\right],
\end{equation}
the same as given in Eq.~\eqref{eq:f_fp}.
This equation can be written as $\partial_t f(x,t)=\partial_x J^{f}(x,t)$, where $J^f$ is the probability flux, a function such that $J^f(x)\Delta$ corresponds to the probability mass flowing from $x+\Delta$ to $x$. To ensure the conservation of probability in $[0;1]$, we impose $J^f=0$ at the boundaries, meaning that no probability mass comes in or goes out during the dynamic evolution of the process. 

In other words, writing $I_f(t)=\int_{0}^{1}\dint x~f(x,t)$, direct integration of Eq.~\eqref{eq:appendix_fp} leads to $\dot{I}_f(t)=J^f(1,t)-J^f(0,t)=0$, ensuring that $I_f(t)=1$ at all times.
Keeping the next term of order $\Delta$ only slightly alters the equation:
\begin{equation}\label{eq:app_fp_delta}
\partial_t f(x,t) = -\epsilon\partial_x\left[(1-2x)f(x,t)\right]+ \partial_{xx}\left[\left(\mu x(1-x)+\epsilon\Delta\right)f(x,t)\right].
\end{equation}
Recalling now that a Fokker-Planck equation of the form
\begin{equation}\label{eq:generic_fp}
    \partial_t p(y,t) = - \partial_y\left[a(y,t)p(y,t)\right]+\partial_{yy}\left[b(y,t)p(y,t)\right]
\end{equation}
corresponds to the It\^o stochastic differential equation
\begin{equation}
    \dot{y} = a(y,t)+\sqrt{b(y,t)}\eta(t)
\end{equation}
where $\eta$ is a brownian white noise, one readily recovers Eq.~\eqref{eq:sde_x}.
Physically, the $0$-flux boundary condition corresponds to a reflecting boundary condition: a ``wall'' that prevents $x$ from getting out of $[0;1]$.

\subsection*{Determining the stationary solution}
Looking for a stationary solution, one sets the right-hand side of Eq.~\eqref{eq:appendix_fp} to $0$, looking to solve
\begin{equation}
    \frac{f'_0(x)}{f_0(x)}=\left(\alpha -1\right)\frac{1-2x}{x(1-x)}\quad \text{with}\quad \alpha:= \frac{\varepsilon}{\mu}
\end{equation}
which, after direct integration, yields $f_0(x)\propto \left(x(1-x)\right)^{\alpha-1}$. Integrating for $x\in[0;1]$ allows one to find the normalisation constant  in terms of the Beta function, or equivalently as a ratio of Gamma functions, to get Eq.~\eqref{eq:stat_sol}.

\section{Change of variables under an SDE} 
\label{sec:change_of_variables_under_an_sde}

Obtaining Eq.~\eqref{eq:phi_fp} and understanding the rationale behind the change of variables of Eq.~\eqref{eq:phi_def} is easier by starting from Eq.~\eqref{eq:sde_x}. 

Imposing a change of variables $x\rightarrow \varphi(x)$ leads to a new stochastic differential equation for $\varphi$, which after applying the It\^o rule for differentiation reads
\begin{equation}\label{eq:phi_sde}
\frac{\dint \varphi(x)}{\dint t} = \varepsilon(1-2x)\varphi'(x)+\mu x(1-x)\varphi''(x) +\sqrt{2\mu x(1-x)}\varphi'(x)\eta(t),
\end{equation}
which is still difficult to interpret because of the dependence on $x$ of the term in front of the white noise $\eta$. 

Picking however $\varphi'(x)= \frac{1}{\sqrt{x(1-x)}}$ amounts to $\varphi(x)=\arcsin(2x-1)$ and rids us of this dependence. Computing the derivatives $\varphi'=2/\cos\varphi$ and $\varphi''=-4\tan \varphi / \cos^2 \varphi$ and replacing in Eq.~\eqref{eq:phi_sde}:
\begin{equation}\label{eq:phi_sde2}
\dot{\varphi} =-\left(2\epsilon-\mu\right)\tan \varphi + \sqrt{2\mu}\eta(t), 
\end{equation}
which because of the equivalence between stochastic differential equations and Fokker-Planck equations discussed in Appendix~\ref{eq:appendix_fp} leads to Eq.~\eqref{eq:phi_fp}. As before, imposing the reflecting boundary conditions $J^g(\pm\pi/2,t)=0$ ensures conservation of probability.

Keeping instead the term of order $\Delta$ given in Eq.~\eqref{eq:app_fp_delta} leads first to the Langevin equation
\begin{equation}\label{eq:x_alt_langevin}
\dot{x} = \varepsilon(1-2x)+\sqrt{2\mu x(1-x) + 2\varepsilon\Delta}\eta(t),
\end{equation}
which leads to the change of variables
\begin{equation}\label{eq:phi_alternative}
 \phi = \arctan\left(\frac{2x-1}{2\sqrt{x(1-x)+\alpha \Delta}}\right),
 \end{equation} 
 where now $\abs{\phi}\leq \arctan\left(1/\sqrt{2\alpha\Delta}\right)\approx \frac{\pi}{2}-2\sqrt{\alpha\Delta}$, and naturally one can check that the definition of $\phi$ corresponds to $\varphi$ as $\Delta\to0$, with $\phi\approx \varphi - 2\alpha\Delta \tan(\varphi)$ to leading order in $\Delta$. The analysis in the limit $N\to\infty$ therefore holds only in the limit $\tan(\varphi)\ll \frac{N}{2\alpha}$.

 This new variable actually verifies the very same SDE, Eq.~\eqref{eq:phi_sde}, but with a different boundary.


\section{Schr\"odinger from Fokker-Planck} 
\label{sec:rewriting_as_a_schrodinger}
The following is a common ``trick'' to transform a non-hermitian dynamic evolution coming from a Fokker-Planck equation with drift into a hermitian evolution determined by a Schr\"odinger equation. We start from a generic Fokker-Planck equation such as the one defined in Eq.~\eqref{eq:generic_fp}, but with constant $b(y,t)=1$ and time-independent drift, which we represent with the derivative of some function $A$, $a(y,t)=-A'(y)$. The resulting Fokker-Planck equation reads
\begin{equation}\label{eq:gen_fp_2}
\partial_t p(y,t) = \partial_y \left[A'(y)p(y,t)\right] + \partial_{yy}p(y,t)
\end{equation}
and has a stationary solution that can be written as a Boltzmann distribution $p_0(y)=e^{-A(y)}/Z$, where $Z$ is a constant ensuring normalisation.

We next introduce a function $\Psi$ verifying $p(y,t)=e^{-A(y)/2}/\sqrt{Z}\Psi(y,t)$. We can compute derivatives to find
\begin{equation}\label{eq:psi_derivatives}
\begin{split}
\partial_y \left[A'(y)p(y,t)\right] &= e^{-A(y)/2}/\sqrt{Z}\left[\left(A''(y)-\frac{A'(y)^2}{2}\right)\Psi(y,t)+A'(y)\partial_y \Psi(y,t)\right]\\
\partial_{yy}p(y,t) &= e^{-A(y)/2}/\sqrt{Z}\left[-\frac{1}{2}\left(A''(y)-\frac{A'(y)^2}{2}\right)\Psi(y,t)-A'(y)\partial_y \Psi(y,t)+\partial_{yy}\Psi(y,t)\right].
\end{split}
\end{equation}

Adding these terms and simplifing, we find the following Schr\"odinger's equation for $\Psi$:
\begin{equation}\label{eq:schrodinger_ex}
-\partial_t \Psi(y,t) = \mathbf{H}\Psi,
\end{equation}
where the Hamiltonian is here defined as
\begin{equation}
\mathbf{H} = -\partial_{yy}+V(y),\quad V(y) = -\frac{1}{2}\left(A''(y)-\frac{A'(y)^2}{2}\right).
\end{equation}

Equation~\eqref{eq:schrodinger_psi} simply uses this substituion, with $\int \dint \varphi~ \tan \varphi=\log \cos \varphi$ playing the role of $A(y)$ (up to a multiplicative constant).


\section{Properties of the solution} 
\label{sec:properties_of_the_solution}
We take the solutions in Eq.~\eqref{eq:sol_array} as those given in \cite{tacseli2003exact}. We first check that they satisfy the boundary condition.
\subsection{Checking the boundary condition} 
\label{sub:checking_the_boundary_condition}
We recall that
\begin{equation}\label{eq:hypergeo_def}
~_2F_1\left(-k,a;b, u\right)=\sum_{l=0}^{k} \binom{k}{\ell}(-1)^\ell \frac{\Gamma(a+\ell)}{\Gamma(a)}\frac{\Gamma(b)}{\Gamma(b+\ell)}u^\ell.
\end{equation}

In this case, direct differentiation in Eq.~\eqref{eq:sol_array} for e.g. even modes $n=2k$ in the limit $\varphi\to\pm\frac{\pi}{2}$leads to
\begin{equation}
\begin{split}
\frac{\dint }{\dint \varphi}\left( _2F_1\left(-k,\beta+k;\beta+\frac{1}{2}, \cos^2\varphi\right)\right) &= 2\sin\varphi\cos\varphi\frac{k(\beta+k)}{\beta+1/2}+\mathcal{O}(\cos \varphi)\\
&\approx \pm 2\left(\frac{\pi}{2} \mp \varphi\right)\frac{k(\beta+k)}{\beta+1/2}.
\end{split}
\end{equation}

With this one can directly compute, with $_2F_1\left(-k,\beta+k;\beta+\frac{1}{2},1\right):=c_1$ and for $\varphi\to\pm\frac{\pi}{2}$:
\begin{equation}
\begin{split}
\beta\tan{\varphi}\Psi_{2k}(\varphi)+\Psi_{2k}'(\varphi)&\approx  \beta\tan{\varphi}\cos^\beta\varphi  ~c_1-\beta\tan{\varphi}\cos^{\beta}\varphi~c_1\pm 2\cos^{\beta}\varphi~ \left(\frac{\pi}{2} \mp \varphi\right)\frac{k(\beta+k)}{\beta+1/2}\\
&\approx \pm2\frac{k(\beta+k)}{\beta+1/2}\left(\frac{\pi}{2} \mp \varphi\right)^{1+\beta},
\end{split}
\end{equation}
which after multiplication with $\cos^\beta\varphi\approx\left(\frac{\pi}{2}\mp\varphi\right)^\beta$ proves Eq.~\eqref{eq:equivalence} for $n=2k$. The proof for odd $n=2k+1$ is strictly equivalent. It therefore follows that the solutions of \cite{tacseli2003exact}, although found initially for vanishing boundary conditions, also satisfy the boundary condition given in Eq.~\eqref{eq:schrodinger_bc}.


\subsection{Explicit expressions} 
\label{sub:explicit_expressions}
In this section we discuss the explicit expressions of the functions $f_n$ and the constants $A_n$.

The constants $A_n$ are set so that $\int_{-\pi/2}^{\pi/2}\Psi_n\Psi_m=\delta_{n,m}$, and therefore implies, in terms of the variable $\alpha$,

\begin{equation}\label{eq:a_ksdef}
\begin{split}
A_{2k}(\alpha) &= \left(\int_{-\frac{\pi}{2}}^{\frac{\pi}{2}} \dint \varphi~\cos^{2\alpha - 1}\varphi~_2F_1\left(-k,\alpha+k-\frac{1}{2};\alpha, \cos^2\varphi\right)^2\right)^{-1/2}\\
A_{2k+1}(\alpha)&= \left(\int_{-\frac{\pi}{2}}^{\frac{\pi}{2}} \dint \varphi~\cos^{2\alpha - 1}\varphi~ \sin^2\varphi~_2F_1\left(-k,\alpha+k+\frac{1}{2};\alpha, \cos^2\varphi\right)^2\right)^{-1/2}.
\end{split}
\end{equation}

To substitute and find the expressions of $f_n(x)$, we recall that
\begin{equation}
\sin \varphi = 2x-1,\quad \cos\varphi = 2\sqrt{x(1-x)}
\end{equation}
and get, using Eq.~\eqref{eq:sol_array} and replacing into $f_n(x) = \frac{\sqrt{g_0(\varphi(x))}\Psi_n(\varphi(x))}{2\sqrt{x(1-x)}}$, the explicit expression

\begin{align}
f_{2k}(x) &= A_{2k}(\alpha)\sqrt{\frac{\Gamma(\alpha+1/2)}{\sqrt{\pi}\Gamma(\alpha)}}~_2F_1\left(-k,\alpha+k-\frac{1}{2};\alpha, 4x(1-x) \right)(4x(1-x))^{\alpha -1}\\
f_{2k+1}(x)&= A_{2k+1}(\alpha)\sqrt{\frac{\Gamma(\alpha+1/2)}{\sqrt{\pi}\Gamma(\alpha)}}~_2F_1\left(-k,\alpha+k+\frac{1}{2};\alpha, 4x(1-x)\right)(4x(1-x))^{\alpha-1}(2x-1).
\end{align}


\subsection{Computing the moments of the distribution}
\label{ssec:moments}
To understand the dynamics of the moments of the distribution
\begin{equation}
    \mathbb{E}\left[x^m(t)\right]=\int_{0}^{1}\dint x~f(x,t)x^m
\end{equation}
it is necessary to understand the behaviour of $B_{n,m}=\int_{0}^{1}\dint x~f_n(x)x^m$. Owing to the parity of $f_n(x)$ with respect to $x=1/2$ it is clear that for even moments $m=2p$ only even modes $n=2k$ will be non zero, and vice versa for odd moments and modes.

We therefore develop the computation of even moments only, as the extension to odd moments is direct. We wish to evaluate the integral $\int_{0}^1 \dint x~f_{2k}(x)x^{2p}=2\int_{0}^{1/2} \dint x~f_{2k}(x)x^{2p}$, after changing variables as $t=4x(1-x)$, it is clear that this integral is proportional to
\begin{equation}
 I_{2p,2m}=\int_{0}^{1}\dint t~_2F_1\left(-k,\alpha+k-\frac{1}{2};\alpha, t\right)t^{\alpha-1}(1-t)^{p-1/2}.
\end{equation}

After expanding the hypergeometric function and integrating explicitly, we find
\begin{equation}
I_{2k,2p} =\frac{\Gamma(\alpha)\Gamma(1/2+p)}{\Gamma(\beta+k)}\sum_{l=0}^{k}\binom{k}{l}(-1)^l \frac{\Gamma(\beta+k+l)}{\Gamma(\beta+1+l+p)} = \frac{\Gamma(\alpha)\Gamma(1/2+p)}{\Gamma(\beta+k)} S_{2k,2p}
\end{equation}
requiring then the explicit computation of the sum $S_{2k,2p}$. 

Mind that for $m=2p+1$ the only modes that contribute are $n=2k+1$, and the equivalent of the previous integral is

\begin{equation}
    I_{2k+1,2p+1} = \frac{\Gamma(\alpha)\Gamma(3/2+p)}{\Gamma(\beta+k+1)}\sum_{l=0}^{k}\binom{k}{l}(-1)^l \frac{\Gamma(\beta+1+k+l)}{\Gamma(\beta+2+l+p)} = \frac{\Gamma(\alpha)\Gamma(3/2+p)}{\Gamma(\beta+k+1)} S_{2k+1,2p+1}
\end{equation}

We discuss this for $k\geq 1$ in two situations, $k>m$ and $k\leq m$.

\subsubsection{First case: $k>m$} 
\label{ssub:first_case_}
We can then write the sum $S_{2k,2p}$ as
\begin{equation}
\sum_{l=0}^{k}\binom{k}{l}(-1)^l \prod_{i = p+1}^{k-1} (\beta + l +i),
\end{equation}
which, written as such, leads us to introduce the function
\begin{equation}\label{eq:polynomial}
    P(X) = \sum_{l=0}^{k}\binom{k}{l}(-1)^l X^{\beta + l + k-1}=X^{\beta + k - 1} (1 - X)^k .
\end{equation}

Applying the generalized Leibniz rule to compute the $k-p-1$-th derivative of this function, we obtain directly that $S_{2k,2p}= P^{(k - p - 1)}(1)=0$ in this case. A similar calculation can be done for $S_{2k+1,2p+1}$, and it follows therefore that
\begin{equation}
    \int_{0}^{1}\dint x~f_n(x)x^m =0\quad \text{ for } n>m. 
\end{equation}

\subsubsection{Second case: $k\leq m$} 
\label{ssub:second_case_}
In this case, we now write the sum as
\begin{equation}
\sum_{l=0}^{k}\binom{k}{l}(-1)^l \prod_{i = k}^{p} \frac{1}{\beta + l +i},
\end{equation}
which can instead be seen as the result of successive integrations on the function defined in Eq.~\eqref{eq:polynomial}.

To compute it, we define the functions $\betaf_0(t;a,b) = t^{a-1} (1-t)^{b-1}$ and $ \betaf_{n+1}(t;a,b) = \int_0^t\dint u~ \betaf_n(u;a,b)$, with $\betaf_1$ corresponding to the standard incomplete Beta function. With this definition, the sum reads

\begin{equation}
    S_{2k,2p} = \int_0^1\dint u~ \betaf_{p-k}(u;\beta + k + 1,k+1),
\end{equation}
while on the other hand successive integration by parts gives
\begin{equation}
\begin{split}
\betaf_n(1;a,b) &= \left[\sum_{j=0}^{n-1} (-1)^{j+1} \frac{(t-1)^{j+1}}{\Gamma(j+2)} \betaf_{n-j}(u;a,b) \right]_0^1 + (-1)^{n} \int_0^1\dint u~ \frac{(t-1)^{n}}{\Gamma(n+1)} \betaf_0(u;a,b) \\
&=\frac{\betaf(n+a,b)}{\Gamma(n+1)}.
\end{split}
\end{equation}

Finally, gathering everything we get

\begin{equation}
S_{2k,2p} = \frac{\Gamma(p + \beta + 1)\Gamma(k+1)}{\Gamma(p+\beta+k+2)\Gamma(p-k+1)},
\end{equation}
while replacing $\beta\to \beta+1$ gives the similar expression
\begin{equation}
    S_{2k+1,2p+1} = \frac{\Gamma(p + \beta + 2)\Gamma(k+1)}{\Gamma(p+\beta+k+3)\Gamma(p-k+1)}.
\end{equation}

The final result follows,

\begin{equation}\label{eq:momentsf}
B_{n,m}=\int_{0}^1 \dint x~f_n(x)x^m = A_n(\alpha)\sqrt{\frac{\Gamma(\alpha+1/2)}{\sqrt{\pi}\Gamma(\alpha)}}I_{n,m} \mathbf{1}\left(n\leq m\right)
\end{equation}

with
\begin{equation}\label{eq:inm}
\begin{split}
I_{2k,2p+1} &= 0\\
I_{2k,2p} &= \frac{\Gamma(\alpha)\Gamma(1/2+p)\Gamma(p + \beta + 1)\Gamma(k+1)}{\Gamma(\beta+k)\Gamma(p+\beta+k+2)\Gamma(p-k+1)}\\
I_{2k+1,2p+1} &= \frac{\Gamma(\alpha)\Gamma(3/2+p)\Gamma(p + \beta + 2)\Gamma(k+1)}{\Gamma(\beta+k+1)\Gamma(p+\beta+k+3)\Gamma(p-k+1)}, 
\end{split}
\end{equation}
allowing then for explicit computation of the dynamics of $\mathbb{E}\left[x^m(t)\right]$.

\section{Stochastic calculus techniques}
\label{sec:stoch}
In this Appendix, we shall directly integrate stochastic differential equations describing the model to obtain information on the covariances of moments $x^n(t)$. We begin by looking at the covariance $\text{Cov}(x(t+T),x(T))$.

A direct integration of Eq.~\eqref{eq:sde_x} leads to
\begin{equation}\label{eq:integrated_sde_x}
x(t+T) = x(T) + \varepsilon t - 2\varepsilon\int_{T}^{t+T}\dint s~x(s)+\int_{T}^{t+T}\dint s~\sqrt{2\mu x(s)(1-x(s))}\eta(s).
\end{equation}

Taking now the covariance with $x(t)$ and using linearity,
\begin{equation}
\begin{split}
\text{Cov}(x(t+T),x(T)) =& \text{Cov}(x(T),x(T))-2\varepsilon\int_{T}^{t+T}\dint s~\text{Cov}(x(s),x(T))\\
&+\int_{T}^{t+T}\dint s~\mathbb{E}\left[\sqrt{2\mu x(s)(1-x(s))}x(T)\eta(s)\right]
\end{split}
\end{equation}
with the last integral being equal to $0$, as
\begin{equation}
\mathbb{E}\left[\sqrt{2\mu x(s)(1-x(s))}x(T)\eta(s)\right]=\mathbb{E}\left[\sqrt{2\mu x(s)(1-x(s))}x(T)\right]\mathbb{E}\left[\eta(s)\right]=0.
\end{equation} 

Taking finally the derivative with respect to $t$ and solving the resulting differential equation we find
\begin{equation}
\begin{split}
\frac{\dint }{\dint s}\text{Cov}(x(T+s),x(T)) &= -2\varepsilon\text{Cov}(x(T+s),x(T))\\
\text{Cov}(x(t+T),x(T)) &\propto e^{-2\varepsilon t}.
\end{split}
\end{equation}
Similarly, one can derive the stochastic differential equation followed by $\sigma_2(x)=x(1-x)$ using the differentiation rule exemplified in Eq.~\eqref{eq:phi_sde}, namely

\begin{equation}
\frac{\dint [x(1-x)]}{\dint t}= \varepsilon - (4\varepsilon+2\mu)x(1-x)+\sqrt{2\mu x(1-x)}(1-2x)\eta(t),
\end{equation}
and as before, we can take the covariance $\text{Cov}(\sigma_2(x(t+T)),\sigma_2(x(T)))$, differentiate with respect to $t$ and find that it satisfies a differential equation, which after integrating reads
\begin{equation}
\text{Cov}(\sigma_2(x(t+T)),\sigma_2(x(T))) \propto e^{-(4\varepsilon +2\mu)t}.
\end{equation}

This method can be extended to computing $C_{n,k}(t+T,T)=\text{Cov}\left(x(t+T)^n, x(T)^k\right)$. Applying It\^o calculus as before, one can show that these functions satisfy the following ODE system:

\begin{equation}\label{eq:ode_syst}
\frac{\dint}{\dint s} \left[C_{n,k}(T+s,T)\right] = - \mu \mathcal{E}_n C_{n,k}(T+s,T)+\mu n(n-1+\alpha)C_{n-2,k}(T+s,T).
\end{equation}

Owing to its triangular structure, it can be diagonalized iteratively to find functions $\sigma_n$, such that $\sigma_n(x)$ is a polynomial of degree $n$ and that the covariances $C_{\sigma_n}(T+s,T)=\text{Cov}\left[\sigma_n(x(T+s)), \sigma_n(x(T))\right]$ satisfy
\begin{equation}\label{eq:ode_diag}
\frac{\dint }{\dint s}C_{\sigma_n}(T+s,T) = -\mu\mathcal{E}_n C_{\sigma_n}(T+s,T).
\end{equation}

Knowing that $\sigma_1(x)=x$ and $\sigma_2(x)=x(1-x)$, it is possible to find the third combination $\sigma_3(x)=(2x-1)\left[\left(1+\frac{2\alpha}{3}\right)(2x-1)^2 -1\right]$. Integrating the equations in Eq.~\eqref{eq:ode_diag}, one finds then that

\begin{equation}
C_{\sigma_n}(t+T,T)\propto e^{-\mu\mathcal{E}_n t}.
\end{equation}

These results can also be obtained directly from the eigenvalues and eigenfunctions of the Schr\"odinger problem.

\end{document}